# Raman analysis of the dehydrogenation process of hydrogenated monolayer graphene

Tom Fournier,[1,2] Kelvin Cruz,[3] Marc Monthioux,[1] Benjamin Lassagne,[2] Lionel Petit,[2] Sébastien Moyano,[1] Pascal Puech,[1*] Fabrice Piazza[3*]

[1]*Centre d'Elaboration des Matériaux et d'Etudes Structurales (CEMES), UPR8011 CNRS, Université de Toulouse, 29 Rue Jeanne Marvig, 31055 Toulouse cedex 4, Toulouse, France*
[2]*Laboratoire de Physique et Chimie des Nano-Objets (LPCNO), UMR5215 CNRS, INSA, Université de Toulouse, 135 avenue de Rangueil, 31077 Toulouse cedex 4, Toulouse, France*
[3]*Laboratorio de Nanociencia, Pontificia Universidad Católica Madre Y Maestra, Autopista Duarte km 1 1/2, Apartado Postal 822, Santiago de Los Caballeros, Dominican Republic*
*Corresponding authors: Pascal Puech pascal.puech@cemes.fr and Fabrice Piazza fpiazza@pucmm.edu.do

**Abstract**

Creating defects in graphene by hydrogenation, either to achieve hydrogen chemisorption or partial etching, is a way to open an electronic band gap in graphene. Understanding the range of stability conditions of partially etched or hydrogenated graphene is crucial for application, as processing conditions (*e.g.* temperature) and quality control (characterization) conditions may result in modifying the material through partial or full dehydrogenation, and subsequent alteration of its electronic properties. This work reports a study of various dehydrogenation conditions of hydrogenated or hydrogen-etched monolayer graphene (1LG), either free-standing or supported by an interferential (SiO$_2$/Si) substrate, using incremental annealing under nitrogen atmosphere up to 400 °C. Materials were investigated by Raman spectroscopy. Indeed, it has been known since 2012 that the intensity ratio of two Raman bands activated by double resonance, D over D' ($I_D/I_{D'}$) can be used to identify the type of defects in defective graphene. It is shown that hydrogenated 1LG, characterized by a large $I_D/I_{D'}$ ratio (~9-15), is stable provided annealing remains below 300 °C. On the other hand, defective 1LG resulting from hydrogen etching remains stable up to 400 °C, whether the 1LG is hydrogenated on one side or both sides, while a modification in the type and proportions of defects is likely. Experimental conditions for the safe use of Raman spectroscopy, otherwise able to induce specimen overheating because of the laser energy and power, are also determined and discussed.

## 1. Introduction

Tailoring complex electronic properties of mono/bilayer graphene (1LG and 2LG, respectively) is already possible by applying an external electric field [1]. Another way to do so is to use chemical functionalization, such as hydrogenation or fluorination, but this is still under progress. Regarding hydrogenation, some primary results were reported in 2009 [2] and others more recently [3,4]. A recent review was published by K.E. Whitener [5]. In this regard, the main difficulty is to achieve locally a high hydrogenation rate for opening an electronic band gap in the graphene electronic structure. Even if the published work shows some successful hydrogenation, the possibility or risk of etching, *i.e.* removing carbon atoms from the graphene layer, is not considered enough. Indeed, by changing the experimental parameters of the hydrogenation setup, it is possible to either clean, functionalize, or etch the graphene layers [6]. The conditions of two-side hydrogenation of 1LG and 2LG are not similar [7] and are also different for one-side hydrogenation of 1LG (work in progress). Hydrogen chemisorption on the graphene surface can be switched to carbon atom etching if the hydrogenation conditions are harshened (according to behavior reported in the work by Delfour *et al* [8]).

Raman spectroscopy is a major tool for studying graphenes [9]. Since the work by Eckmann *et al* [10], it is known that the intensity ratio $I_D/I_{D'}$ of the D and D' Raman bands relates to the type of defects in graphene at stage 1 of defectiveness (roughly corresponding to the intensity ratio of the D band over the G band, $I_D/I_G < 4$ at 515 nm). An $I_D/I_{D'}$ ratio value around 13 corresponds to hydrogenation or fluorination (hence to the occurrence of $sp^3$C atoms), $I_D/I_{D'}$ around 7 indicates vacancies, and $I_D/I_{D'}$ around 3.5 relates to the contribution of graphene edges. Theoretical work addressed the reasons for the variation of the relative intensity between the two bands and confirmed the $I_D/I_{D'}$ ratio as a reliable parameter to probe the type of defects [11]. However, if the number of defects is large ($I_D/I_G > 4$ at 515 nm, corresponding to stage 2 of defectiveness [10]), this ratio does not show a clear dependence on the type of defect.

In the literature, it is not so clear at which annealing temperature chemisorbed hydrogen is removed from the graphene surface. It ranges from 150°C [12] to 400°C [2], although conditions were not exactly the same. Based on the evolution of the $I_D/I_G$ ratio, Luo *et al* [13] suggested a mechanism involving two desorption thresholds, one at 100-150 °C and another at 260-300 °C. They are based on the energy barriers calculated for desorbing hydrogen dimers respectively chemisorbed in para and ortho positions onto a graphite surface, which both are larger than for the meta position [14]. For genuine graphane or diamane (a single or double layer of $sp^3$C atoms, respectively, hydrogenated on both sides, with the diamond or lonsdaleite structure in the case of diamane), temperatures beyond this range are expected to achieve hydrogen desorption (4 h annealing at 600 °C in vacuum for bulk graphane [15]). Many calculations dealing with the transformation of graphene into periodic $sp^3$C structures upon hydrogenation were reported, using a thermodynamical model to account for strain [16], but also considering a first principle model [17] to account for structure stability, as well as molecular dynamics [18] to analyze the diamane formation.

In this work, we investigate the annealing of hydrogenated 1LG to achieve full dehydrogenation. The resulting samples are characterized by Raman scattering. The parameters of the study include the type of substrate (none, *i.e.* suspended, or SiO$_2$/Si) supporting the 1LG specimens. We also discuss the experimental conditions for safely carrying out Raman characterization on such materials, as laser illumination was found to possibly affect the hydrogen coverage [12].

## 2. Experimental setup and samples

1LG grown by chemical vapor deposition (CVD) lying onto transmission electron microscopy (TEM) grids were purchased from *Graphenea™*. The 1LG was supported by a thin (12 nm) holey amorphous carbon film exhibiting a periodic array of 2 μm-large holes, itself supported by an Au grid. In that case, then, 2 μm-large areas of the 1LG are suspended across the holes in the carbon film. Those samples will be called "free-standing-1LG" thereafter. On the other hand, graphene flakes were obtained from a highly orientated pyrolytic graphite (HOPG) piece, and subsequently transferred onto marked SiO$_2$ (285 nm)/Si interferential substrates by the scotch-tape technique [19]. 1LG were selected optically among the flakes from their light absorption and then confirmed by Raman spectroscopy. In the case of 1LG, the full width at half maximum (FWHM) of the 2D band is below 30 cm$^{-1}$ [9] (also see the Supporting Information). Then, parts of each 1LG flake investigated were masked with a polymethylmethacrylate (PMMA) 100 nm-thick resist layer so that the subsequent hydrogenation affects only ribbon-like areas. The PMMA polymer was removed using acetone and isopropyl alcohol before annealing and subsequent Raman characterization. This masking procedure was used to easily identify the region of interest which, after hydrogenation becomes difficult to observe due to dramatic decrease in optical contrast. Those samples will be called "supported-1LG" thereafter.

Hydrogenation and hydrogen etching were carried out in a hot-filament reactor in which molecular hydrogen is dissociated onto resistively-heated surface-carburized tungsten filaments located 4.4 cm away from the substrate holder [7]. Hydrogenation conditions were a 1700 °C filament temperature, with 10 Torr $H_2$ residual pressure and an $H_2$ flow of 1 sccm during 60 minutes for the free-standing 1LG (positioned vertically on the substrate holder), and a 2000 °C filament temperature, with 85 Torr $H_2$ residual pressure and an $H_2$ flow of 50 sccm during 60 minutes for the supported-1LG (positioned horizontally). It is expected that the free-standing 1LG will be hydrogenated on both sides, whereas the supported-1LG will be hydrogenated on one side only.

To achieve hydrogen etching and obtain defective 1LG, the free-standing 1LG was subjected to another set of conditions: 1900°C filament temperature, residual pressure of 50 Torr, 10 sccm $H_2$ flow.

In our experimental setup, the temperature of the specimen is difficult to estimate.

Ultra-low-voltage (5 kV) TEM was used to observe the morphology of the specimens while avoiding dehydrogenation (as the knock-on energy for hydrogen atoms is higher than 5 keV).

To anneal the samples and investigate the subsequent dehydrogenation process, a Linkam cryostat/heater (-195 to 600 °C) in ultra-high purity $N_2$ at atmospheric pressure with water cooling was used. The samples were heated incrementally from 100 to 300 °C (supported-1LG) or to 400 °C (free-standing 1LG) as follows: once the cryostat was closed, the specimen was heated up to the first set temperature (*i.e.* 100 °C) with a 5 min dwell, then cooled down to 25 °C, and characterized by Raman spectroscopy with the cryostat opened. Then the cryostat was closed again, and the specimen was heated up to the second set temperature (*e.g.* 200 °C) with a 5 min dwell, and so on. Both heating and cooling rates were 50 °C/min. For free-standing 1LG, the temperature increment was 100 °C, whereas, for supported-1LG, the temperature increment was 20 °C from 100 to 200 °C, and then 50 °C from 200 to 300 °C. This specific heat-treatment procedure was used instead of carrying out the Raman while the specimens were heated up because it was observed that cumulating the laser energy with the thermal energy from the annealing procedure resulted in the material modification and variation of the relative intensity of the D band.

Raman spectra were acquired in air using a x100 magnification objective lens (numerical aperture of 0.9; spot size of around 1 $\mu m^2$) and *(i)* red excitation (633 nm) with a laser power of 0.5 µW (obtained by applying a 3.2 % filter to the full laser power of 17 mW) during 100 s in the case of the free-standing 1LG to avoid laser-induced heating, or *(ii)* green excitation (532 nm) with a laser power of 1 mW (obtained with a 1 % filter applied to the full laser power of 100 mW) during 100 s in the case of the supported-1LG. The laser power values were measured after the objective lens, hence they correspond to the power actually received by the specimens. For the studies of the effects of the laser power on the structure, 1, 3.7, 10, and 25 % (50 % at 633 nm) filters were used for durations between 1 to 60 minutes with both wavelengths on supported-1LG. As both wavelengths generated similar results, only those at 633 nm will be shown.

The Raman spectra of the starting 1LG materials (Supporting Information) exhibit only the G band and the typically narrow 2D band (FWHM in the range 25-26 cm$^{-1}$), and no D band.

## 3. Results

### 3.1. Free-standing 1LG

We selected two kinds of samples, one subjected to etching conditions, and another subjected to hydrogenation conditions. The former was found to exhibit an $I_D/I_{D'}$ ratio close to 4 and the latter was found to exhibit an $I_D/I_{D'}$ ratio in the range of 10. However, regarding the latter, the uncertainty is large due to the low signal/noise ratio and the low intensity of the D' band, hence the $I_D/I_{D'}$ ratio varies from 9 to 15 in all our hydrogenated samples, which confirms the presence of *sp*$^3$-C as defects.

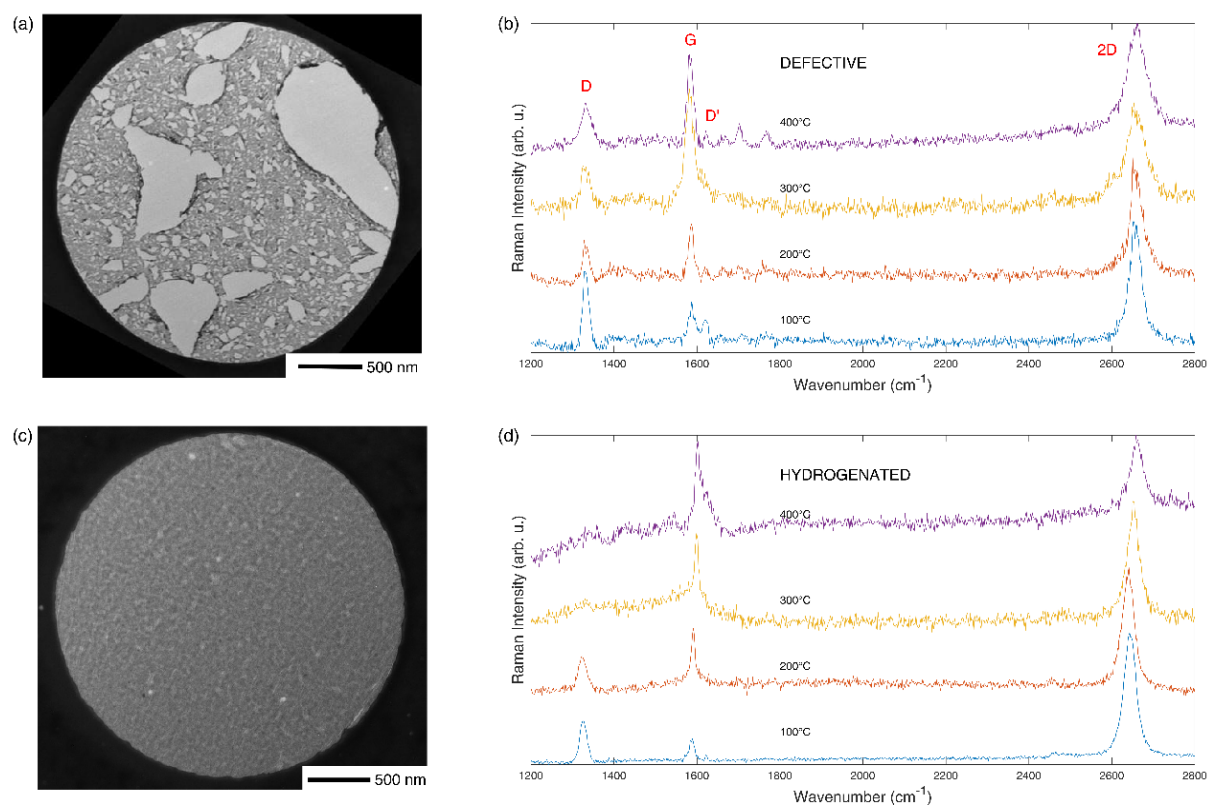

*Figure 1:* Low-magnification TEM images of free-standing 1LG across a hole in the amorphous carbon film (onto a TEM grid). *(a)* after hydrogen-promoted etching and *(c)* after hydrogenation. *(b)* and *(d)*: Raman spectra (at 633 nm) of the samples corresponding to *(a)* and *(c)*, respectively, obtained after increasing incrementally the annealing temperature on the same specimen.

In **Figure 1a** is reported the TEM image of the hydrogen-etched sample. The 1LG film has cracked. The dark grey contrast is due to the widespread presence of PMMA residue already present in the 1LG material as received from *Graphenea™* . Areas exhibiting a clear contrast correspond to the absence of material. In **Figure 1b** is reported the associated Raman spectra at various incremental annealing temperatures in which the undifferentiated background, provided by the PMMA remnants which gradually turned into amorphous carbon upon heat treatment, was removed. Due to the specific experimental procedure described in Section 2, the location on the grid for each spectrum could not be exactly the same. The spectrum at 100 °C is similar to that obtained right after etching treatment. A relatively large and intense D band is observed and remains visible at all the annealing temperatures up to 400 °C. The $I_D/I_{D'}$ ratio value remains in the range of 4 which is consistent with the occurrence of edges as a consequence of the etching effect.

In **Figure 1c** is reported the TEM image of the 1LG material after successful hydrogenation, showing that the continuity of the graphene layer, yet hydrogenated, is preserved. Again, dark grey contrasts correspond to PMMA residues. Light grey contrasts in-between correspond to the hydrogenated 1LG layer underneath. The Raman spectra of the hydrogenated free-standing 1LG after increasing annealing of the material up to 400 °C are reported in **Figure 1d**. A D band occurs after hydrogenation and then vanishes from an annealing temperature of 300 °C, revealing full dehydrogenation and recovering of the original graphene state ($sp^2$-C).

### 3.2. Supported-1LG

In **Figure 2a** is reported an optical image of a 1LG flake, onto which both the protected (masked) and the hydrogenated areas are indicated. A decrease in the optical contrast can be observed for the area exposed to hydrogen.

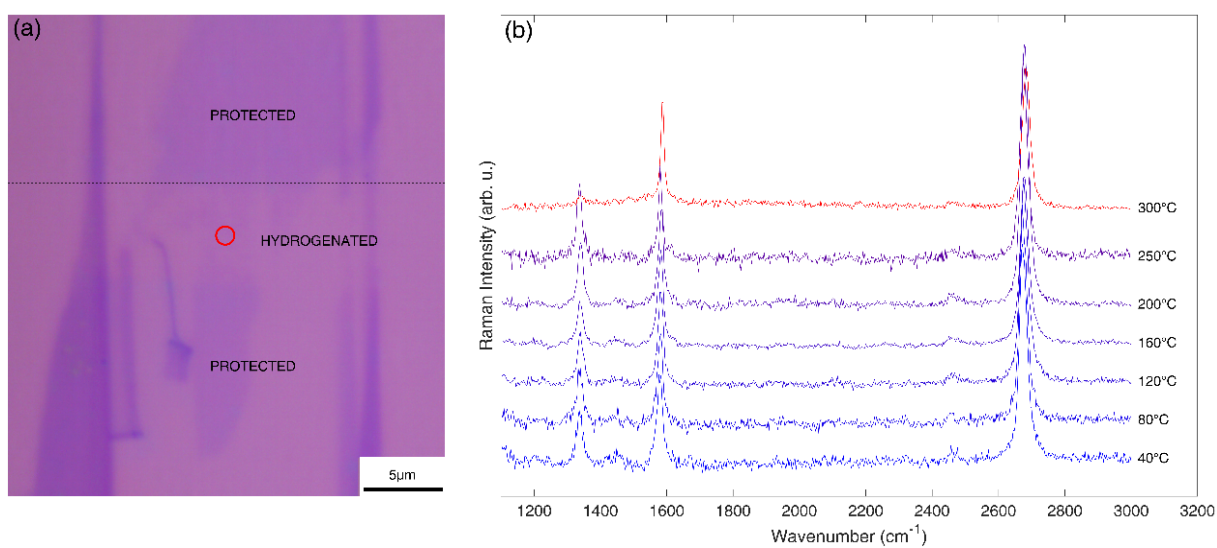

*Figure 2: (a)* Optical image of a 1LG flake on Si/SiO$_2$ substrate exposed locally to hydrogen radicals and *(b)* Raman spectra obtained on the hydrogenated area after annealing at temperatures incrementally increased up to 300°C. The red circle on Figure 2a indicates the localization of the laser spot during Raman measurements.

**Figure 2b** shows the Raman spectra obtained on the hydrogenated area after annealing. The relative intensity of the D' band is very low, making the fitting of the spectra ambiguous. The D band appears after hydrogenation and its relative intensity remains significant up to an annealing temperature of 250°C*.* From 300°C, the relative intensity of the D band becomes very low or vanishes as for the free-standing hydrogenated 1LG (**Fig.1d**). Typically, $I_D/I_{D'}$ ratio values ranging from 10 to 15 are obtained all along the temperature range explored, consistently with the hydrogenated state of the carbon layer. Using 400 °C annealing temperature allowed obtaining spectra (not shown) similar to that obtained by Elias *et al* [2], where doping is present, revealed by an upshift of both the G and 2D bands, and which is already visible at 300 °C in **Fig.2b**. The comparison of the spectra obtained after an annealing at 250°C and 300°C reveals that both the G and 2D bands are upshifted. The shift of the G and 2D bands are $\Delta\omega_G$ = +6.5 cm$^{-1}$ and $\Delta\omega_{2D}$ =+7.9 cm$^{-1}$, respectively, leading to a ratio $\Delta\omega_G / \Delta\omega_{2D}$ of 0.8. In the case of pristine graphene, it was proposed that the ratio $\Delta\omega_G / \Delta\omega_{2D}$ can be used to interpret the origin of the shift of the bands [20]: strain, if the ratio is 2.2; hole doping, if it is 0.7; a combination of strain and hole doping if the value of the ratio is between those two values. This analysis indicates that in the present case, the major contribution is doping. Removing H atoms has restored the Dirac cone allowing doping effect [21].

### 3.3. Laser annealing

To check the dependence of the Raman spectrum features on the laser power, a series of Raman spectra were acquired on the same area every minute up to 60 minutes for a 633 nm wavelength on a hydrogenated supported-1LG exhibiting an $I_D/I_{D'}$ ratio of about 9. The power was started at 0.15 mW to acquire Raman spectra for the first time series, and then the power was successively increased to 0.5, 2, 4, and 8 mW to acquire the spectra for the other time series, respectively.

At a laser power lower than 4 mW, no significant effect was observed. In **Figure 3a** is reported the variation of the $I_D/I_G$ ratio in function of the photoirradiation time at 4 mW. The $I_D/I_G$ intensity decreases with increasing photoirradiation time, but the wavenumber and linewidth of the G band do not change, whereas a strong downshift was expected with increasing heating [22]. This indicates that the heating generated by the laser induces two phenomena, the effects of which on the bond vibrations are opposed and compensate, for instance the combination of heating on the one hand, and of doping by photo-carriers on the other hand. In **Figure 3b** are reported Raman spectra averaged over the first and last 10 acquisitions at the beginning (red) and at the end (blue) of the increasing irradiation experiment, respectively.

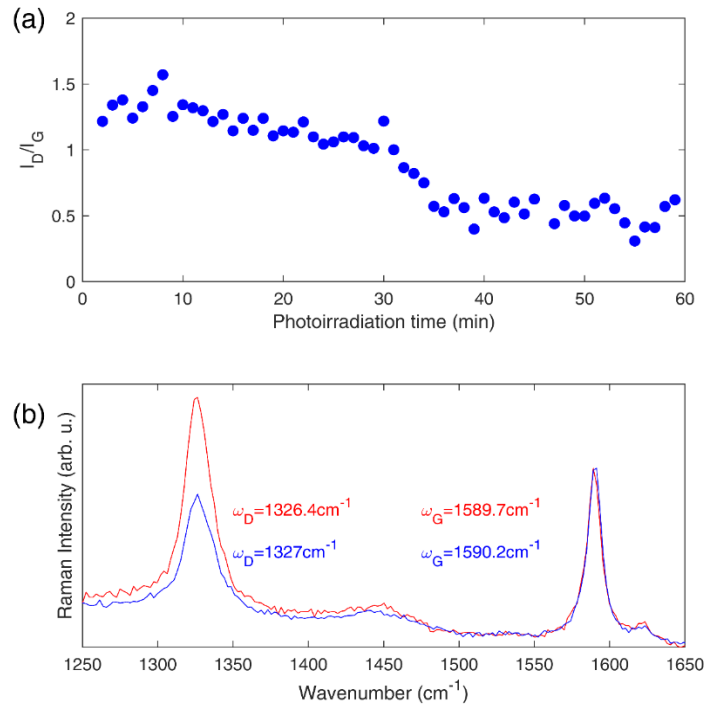

*Figure 3:* Hydrogenated supported-1LG subjected to photoirradiation with a 4 mW 633 nm laser for increasing times. *(a)* $I_D/I_G$ ratio versus photoirradiation time. *(b)* averaged (over 10 spectra) Raman spectra at the begining (red color) and at the end (blue color) of the irradiation process.

After the irradiation series at 4 mW, the $I_D/I_G$ ratio remained constant when using a higher power (8 mW), meaning that the material could not be modified further. The explanation can be that the defects corresponding to hydrogenation were removed and the remaining defects correspond to vacancies and edges. However, it is not ascertained that only dehydrogenation is induced by the laser. For instance, the release of H atoms from the surface is likely to result in reactive carbon atoms able to subsequently oxidize, due to air (oxygen) interaction. As a consequence, oxygenated functions (*e.g.* hydroxyl, epoxy) can be grafted while the carbon atoms involved remain $sp^3$. From the starting time to the end, the $I_D/I_{D'}$ ratio decreases from 15 to 8.5 (see **Figure 3b**).

In fact, increasing the laser power is not a good strategy at 633 nm (or 532 nm). Indeed, hydrogenated graphene is expected to be less thermally conductive than pristine graphene, and thus the heat supplied from the laser leads to a local annealing of the carbon layer. When using a laser power not able to induce overheating, as the desorption in air of the chemisorbed hydrogen occurs at a temperature lower than that of carbon oxidation, the carbon layer is mainly cleaned from its hydrogen content.

## 4. Discussion

Hydrogenation of graphene surface is commonly carried out using processes such as hot filament [7,23] and radio frequency hydrogen plasma [2,24]. However, successfully achieving graphene hydrogenation, hence transforming part or all of the $sp^2$C into $sp^3$C, is not straightforward whatever the method used. Therefore, any claim for the successful synthesis of graphane [2] or diamane [25] should be supported by well-adapted techniques to be fully ascertained and convincing. The main problem with fully hydrogenated carbon layers such as graphane and diamane where all carbon atoms are $sp^3$ and half of them hydrogenated (all of them for graphane) is that the material is transparent for visible light, making interferential substrates useless to locate the hydrogenated material lying on them. With graphene only partly hydrogenated as in this study, the resulting carbon layer is still an optical absorber, meaning that the interferential substrate allows locating the layer to carry out Raman spectroscopy.

In the case of the partly hydrogenated 1LG prepared in this study, dehydrogenation by heating was observed starting from ~300 °C, whether the hydrogenation has affected one side or both sides of the 1LG. We show that hydrogenation is associated with an $I_D/I_{D'}$ ratio larger than 9, which is fully consistent with the measurements by Eckmann *et al* [10]. This allows the occurrence of the D band to be associated with hydrogen chemisorption on carbon atoms as a result of the hydrogenation process. As a matter of fact, tracking the $I_D/I_{D'}$ ratio is enough to ascertain actual hydrogenation instead of etching provided the laser power during the Raman experiment is kept low (*e.g.,* < 4 mW at 633 nm). However, the power value threshold is difficult to estimate as the optical absorbance on defective carbon layers is not identical to that of pristine graphene layer and varies with the type and proportion of defects. Moreover, the actual power value threshold depends on the thermal conductivity of the material, which decreases as the number of defects increases and, thus, increases with the hydrogenation rate [26]. It is worth mentioning that the $I_D/I_{D'}$ ratio is not expected to be sensitive to polarization for $sp^3$-C and for vacancies. As hydrogenation leads to circular activated areas, by symmetry, we do not expect the $I_D/I_{D'}$ ratio to be sensitive to polarization, as predicted theoretically for punctual defects [27]. On the other hand, for hole defects, the sensitivity of the D band to polarization is not excluded due to the existence of two types of edges, zigzag or armchair [28]. This is because the D band is due to an intervalley KK' double resonance process. On the contrary, the D' band, because it results from an intravalley double resonance process, should not be sensitive to polarization. Nevertheless, this effect keeps the $I_D/I_{D'}$ ratio at a moderate value, far from the values characteristic of $sp^3$-C and vacancies.

## 5. Conclusions

We investigated by Raman spectroscopy the thermal stability of partially hydrogenated monolayer graphene, either suspended or supported on a Si/SiO$_2$ interferential substrate. We also considered the case of monolayer graphene partially etched by hydrogen radicals, having a large content of vacancies and/or edges. In the latter case, the Raman $I_D/I_{D'}$ ratio is much lower than in the case of hydrogen chemisorption ($I_D/I_{D'}$ ~9-15), and the defects relate to edges which cannot be cured by annealing. All those hydrogenated or etched graphene materials were obtained by using a hot-filament-assisted hydrogenation process. Full dehydrogenation is achieved following an annealing at 300 °C in ultra-high-purity N$_2$ at atmospheric pressure whereas no structure change is detected in partially-etched graphene. During Raman spectroscopy analysis performed in air, if a 633 nm excitation laser is used, laser power must be kept below 4 mW to avoid any structure modification. Raman spectroscopy is confirmed as a powerful technique to investigate partially-hydrogenated or partially-etched graphene. The $I_D/I_{D'}$ ratio can be used to monitor any structure change such as dehydrogenation or vacancy/edge formation upon processing or characterization.

**Acknowledgements:**
This research is supported by ANR-21-CE09-0003 GLADIATOR project and received exchange funding from CNRS (France) through the French-Dominican International Research Project NEWCA. T.F. thanks NanoX ANR-17-EURE-0009 in the framework of the "Programme des Investissements d'Avenir for the mobility scholarship. This research was partially funded by Fondo Nacional de Innovación y Desarollo Científico y Tecnológico (FONDOCyT), Dominican Republic, grant No. 2018-2019-1A2-087, No. 2020-2021-1A1-066 and No. 2022-1A1-095)

**Contributions:**
Conceptualisation, funding acquisition and project administration: PP and FP.
Investigation, validation, visualisation and writing-review & editing : all authors.
Writing - original draft : MM, BL, PP, and FP.